\documentclass[letter,longauth,traditabstract]{aa}                                    \usepackage{graphicx,longtable,lscape,caption,subcaption}
\usepackage{txfonts}
\usepackage{natbib}
\bibpunct{(}{)}{;}{a}{}{,} 
\begin{document}
\title{The spine of the swan: A \emph{Herschel}\thanks{\emph{Herschel} is an ESA space observatory with science instruments provided by European-led Principal Investigator consortia and with important participation from NASA (Pilbratt et al.~2010).} study of the DR21 ridge and {filaments} in Cygnus~X}

\titlerunning{The spine of the swan}
\authorrunning{M. Hennemann, F. Motte, N. Schneider et al.}

\author{M.~Hennemann\inst{\ref{cea}},
F.~Motte\inst{\ref{cea}},
N.~Schneider\inst{\ref{cea}},
P.~Didelon\inst{\ref{cea}},
T.~Hill\inst{\ref{cea}},
D.~Arzoumanian\inst{\ref{cea}},
S.~Bontemps\inst{\ref{lab},\ref{cnrslab}},
T.~Csengeri\inst{\ref{mpifr}},
Ph.~Andr\'e\inst{\ref{cea}},
V.~Konyves\inst{\ref{cea}},
F.~Louvet\inst{\ref{cea}},
A.~Marston\inst{\ref{hsc}},
A.~Men'shchikov\inst{\ref{cea}},
V.~Minier\inst{\ref{cea}},
Q.~Nguyen~Luong\inst{\ref{cita},\ref{cea}},
P.~Palmeirim\inst{\ref{cea}},
N.~Peretto\inst{\ref{cea}},
M.~Sauvage\inst{\ref{cea}},
A.~Zavagno\inst{\ref{lam}},
L.~D.~Anderson\inst{\ref{wvuni}},
J.-Ph.~Bernard\inst{\ref{toul}},
J.~Di~Francesco\inst{\ref{can}},
D.~Elia\inst{\ref{iaps}},
J.~Z.~Li\inst{\ref{naoc}},
P.~G.~Martin\inst{\ref{cita}},
S.~Molinari\inst{\ref{iaps}},
S.~Pezzuto\inst{\ref{iaps}},
D.~Russeil\inst{\ref{lam}},
K.~L.~J.~Rygl\inst{\ref{iaps}},
E.~Schisano\inst{\ref{iaps}},
L.~Spinoglio\inst{\ref{iaps}},
T.~Sousbie\inst{\ref{iap}},
D.~Ward-Thompson\inst{\ref{ucl}},
G.~J.~White\inst{\ref{openu},\ref{ral}}
}

\institute{AIM Paris-Saclay, CEA/DSM/IRFU -- CNRS/INSU -- Universit\'e Paris Diderot, CEA Saclay, 91191 Gif-sur-Yvette cedex, France
\email{martin.hennemann@cea.fr} \label{cea}
\and 
Universit\'e de Bordeaux, LAB, UMR5804, F-33270 Floirac, France \label{lab}
\and
CNRS, LAB, UMR5804, F-33270 Floirac, France \label{cnrslab}
\and
Max-Planck-Institut f\"ur Radioastronomie, Auf dem H\"ugel 69, 53121 Bonn, Germany \label{mpifr}
\and
Herschel Science Centre, ESAC, ESA, PO Box 78, Villanueva de la Ca\~nada, 28691 Madrid, Spain \label{hsc}
\and
Canadian Institute for Theoretical Astrophysics - CITA, University of Toronto, 60 St. George Street, Toronto, Ontario, M5S 3H8, Canada \label{cita}
\and
Laboratoire d'Astrophysique de Marseille, CNRS/INSU -- Universit\'e de Provence,
13388 Marseille cedex 13, France \label{lam}
\and
Department of Physics, West Virginia University, Morgantown, WV 26506, USA \label{wvuni}
\and
Universit\'e de Toulouse, UPS, CESR, 9 avenue du colonel Roche, 31028 Toulouse Cedex 4, France;
CNRS, UMR5187, 31028 Toulouse, France  \label{toul}
\and
National Research Council of Canada, Herzberg Institute of Astrophysics, 5071 West Saanich Road, Victoria, BC V9E 2E7, Canada \label{can}
\and
Istituto di Astrofisica e Planetologia Spaziali - IAPS, Istituto Nazionale di Astrofisica - INAF,
Via Fosso del Cavaliere 100, I-00133 Roma, Italy \label{iaps}
\and
National Astronomical Observatories, Chinese Academy of
Sciences, A20 Datun Road, Chaoyang District, 100012 Beijing,
China \label{naoc}
\and
Institut d'Astrophysique de Paris, Universit\'e Pierre et Marie Curie (UPMC), CNRS (UMR 7095), 75014 Paris, France \label{iap}
\and
Jeremiah Horrocks Institute, University of Central Lancashire, PR1 2HE, UK \label{ucl}
\and
Department of Physics \& Astronomy, The Open University, Walton Hall, Milton Keynes MK7 6AA, UK \label{openu}
\and
Space Science \& Technology Department, CCLRC Rutherford Appleton Laboratory, Chilton, Didcot, Oxfordshire OX11 0QX, UK \label{ral}
}

\date{Received 2012; accepted 2012}

\abstract
{
In order to characterise the cloud structures responsible for the formation of high-mass stars, we present \emph{Herschel} observations of the DR21 environment.
Maps of the column density and dust temperature unveil the structure of the DR21 ridge and several connected {filaments}.
The ridge has column densities larger than 10$^{23}$\,cm$^{-2}$ over a region of 2.3\,pc$^2$.
It shows substructured column density profiles and branching into two major {filaments} in the north.
The masses in the studied {filaments} range between 130 and 1400\,M$_\odot$ whereas the mass in the ridge is 15000\,M$_\odot$.
{The accretion of these filaments onto the DR21 ridge, suggested by a previous molecular line study, could provide a continuous mass inflow to the ridge.}
In contrast to the striations seen in e.g., the Taurus region, {these filaments} are gravitationally unstable and form cores and protostars.
{These cores} formed in the {filaments} potentially fall into the ridge.
Both inflow and collisions of cores could be important to drive the observed high-mass star formation.
The evolutionary gradient of star formation running from DR21 in the south to the northern branching is traced by decreasing dust temperature.
This evolution and the ridge structure can be explained by two main filamentary components of the ridge that merged first in the south.
}


\maketitle

\begin{figure*}
\centering
\includegraphics[width=\textwidth]{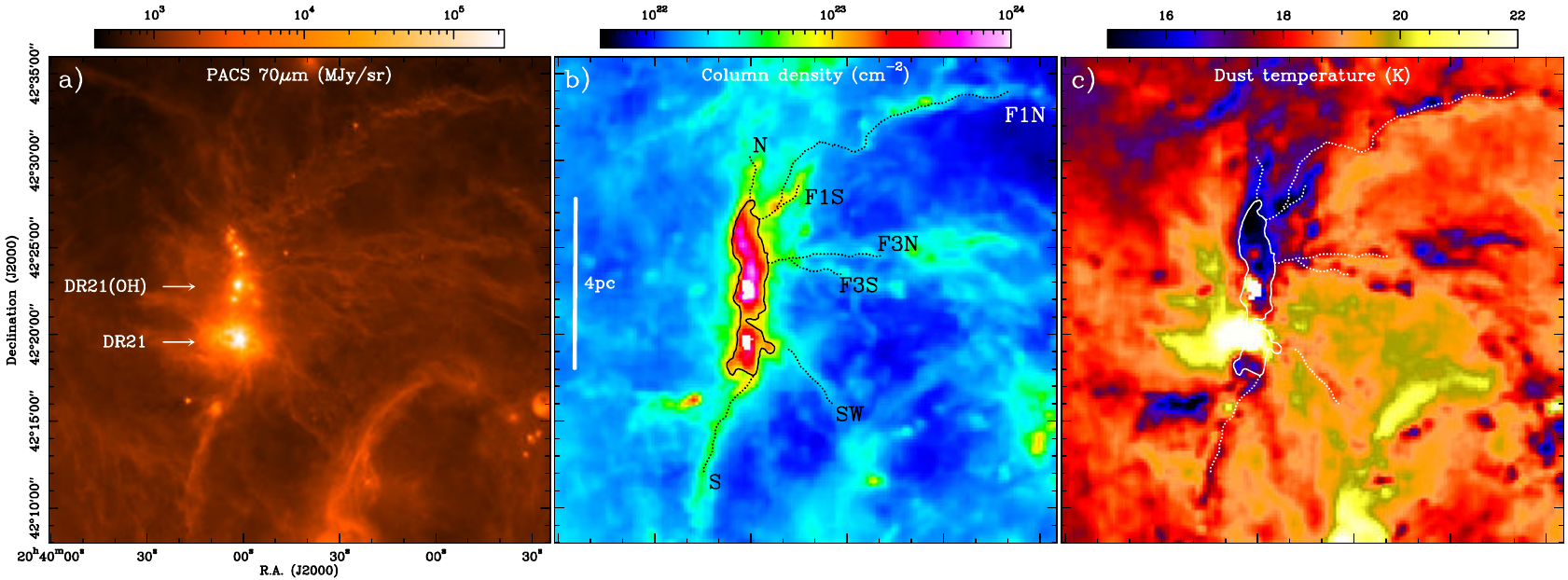}
\caption{\emph{Herschel} maps of the DR21 environment showing {a)} 70\,$\mu$m emission, {b)} column density, and {c)} dust temperature.
The DR21 ridge is delimited roughly by the N$_{\rm H_2} = 10^{23}$\,cm$^{-2}$ contour plotted in panels b) and c).
The {filaments} {selected using \emph{DisPerSE} (see Section~\ref{sec_subfilaments}) are named and} marked with dots along their crests in b) and c).}
\label{fig_cd}
\end{figure*}

\section{Introduction}
The properties of dense structures in interstellar molecular clouds, e.g., the density and temperature distribution, chemical composition, and dynamics, are expected to determine the detailed outcome of the star formation process that varies between an isolated low-mass star and an OB star cluster.
While previous observations have shown that clouds are filamentary \citep[e.g.,][]{bally1987a}, recent \emph{Herschel}\nocite{pilbratt2010a} studies have revealed the ubiquity \citep[e.g.,][]{miville-deschenes2010a,henning2010a} and key importance of filaments for star formation.
Notably, gravitationally unstable filaments are found to fragment into cores \citep{andre2010a}, and filaments generally show narrow central widths of about 0.1\,pc \citep{arzoumanian2011a}.
Resulting self-gravitating cores with sizes of $0.01$--$0.1$\,pc and masses up to $10$\,M$_\odot$ are best candidates to form low- to possibly intermediate-mass stars \citep[see reviews by][]{di-francesco2007a,ward-thompson2007a}.

It is an important question which cloud structures lead to the formation of high-mass stars ($\gtrsim$10\,M$_\odot$) almost exclusively found in stellar clusters or associations \citep[see review by][]{zinnecker2007a}.
The \emph{Herschel} imaging survey of OB Young Stellar objects \citep[HOBYS,][]{motte2010a} observes massive molecular cloud complexes within 3\,kpc distance to probe the cloud environment of OB star-forming cores, their statistical evolution, and the effects of feedback on parental clouds.
Studying the Vela C region, \citet{hill2011a} suggested that ``ridges'', i.e., massive, gravitationally unstable filamentary structures of high column density (\mbox{N$_{\rm H_2}>10^{23}$\,cm$^{-2}$}) that dominate their environment could be preferential sites of high-mass star formation \citep[cf.][]{nguyen-luong2011a}.
The ridge in Vela C shows a complex substructure that could result from stellar feedback \citep{minier2012a}.
Further, intersecting filaments appear to mark sites of stellar cluster formation in the Rosette cloud \citep{schneider2012a}.

This study focusses on the DR21 ridge (also called DR21 filament), the densest and most massive cloud structure in the Cygnus~X region \citep{schneider2006a,motte2007a,roy2011a} at a distance of 1.4\,kpc \citep{rygl2012a}.
It hosts the embedded HII region DR21 \citep[e.g.,][]{roelfsema1989a}, the maser source DR21(OH), and massive protostars \citep{bontemps2010a}.
\citet{schneider2010a} analysed molecular line emission and identified three {``sub-filaments''} F1--F3 that connect to the ridge.
Velocity gradients suggest that material is transported along them towards the ridge, and in the case of F3 a bend and possible direct connection to the DR21(OH) clump is traced \citep[cf.][]{csengeri2011b}.
This behaviour supports the scenario that sustained accretion of inflowing material plays a role in building up massive clumps and cores and setting the stage for high-mass star formation, as suggested by numerical simulations \citep[e.g.,][]{balsara2001a,banerjee2006a,smith2011a}.
Here we exploit the unprecedented sensitivity of \emph{Herschel} far-infrared and submillimeter continuum imaging to trace the detailed column density structure in this region.

\section{\emph{Herschel} observations and data reduction}

Cygnus~X North was observed in the parallel scan map mode with PACS \citep{poglitsch2010a} and SPIRE \citep{griffin2010a} on Dec 18, 2010.
To diminish scanning artifacts, two nearly perpendicular coverages of 2.8$\times$2.8$^\circ$ were obtained in five photometric bands at 70, 160, 250, 350, and 500\,$\mu$m with a scan speed of 20\arcsec/sec.
The data were reduced using the \emph{Herschel Interactive Processing Environment} and the \emph{Scanamorphos} software \citep[][for creating PACS maps]{roussel2012a}.

The 70\,$\mu$m map traces in particular heated dust towards star-forming sites (Fig.~\ref{fig_cd}\,a).
To quantitatively estimate the distribution of the cold dust which generally represents most of the dust mass, we created maps of column density N$_{\rm H_2}$ and dust temperature T$_{\rm d}$ in the way described in \citet{hill2011a,hill2012a}, including offsets to recover the absolute intensity level following \citet{bernard2010a}.
Computed from the 160, 250, and 350\,$\mu$m bands, the maps provide an angular resolution of 25\arcsec (0.17\,pc at 1.4\,kpc).
They have undefined values for pixel groups towards DR21, DR21(OH) {(white pixels in Fig.~\ref{fig_cd}\,b)}, and W75N (see Fig.~\ref{fig_maps}) due to saturation at 250 and 350\,$\mu$m.

\section{The DR21 {filaments}}
\label{sec_subfilaments}

\begin{table*}
\caption{Average properties of the DR21 {filaments} and ridge}
\label{tab_filaments}
\centering
\tiny
\begin{tabular}{lcccccccc}
\hline
  & S & SW & F3S & F3N & F1S & F1N & N & Ridge \\ \hline
Length (pc) & 2.6 & 1.6 & 1.3 & 2.7 & 1.3 & 6.9 & 1.1 & 4.1 \\
Mean crest column density\tablefootmark{a} (10$^{22}$\,cm$^{-2}$) & 4.4 & 1.2 & 1.6 & 2.1 & 7.7 & 1.9 & 4.8 & 41.6 \\
Integration radius (pc) & 0.58 & 0.41 & 0.24 & 0.24 & 0.48 & 0.39 & 0.49 & 1.15 \\
Total mass (M$_{\odot}$) & 1350 & 210 & 130 & 400 & 960 & 1210 & 610 & 15210 \\
Mean line mass (M$_{\odot}$/pc) & 500 & 120 & 100 & 140 & 700 & 170 & 520 & 3670 \\
Weighted mean central width\tablefootmark{b} (pc) & 0.28$\pm$0.07 & 0.31$\pm$0.07 & 0.27$\pm$0.08 & 0.32$\pm$0.11 & 0.26$\pm$0.07 & 0.33$\pm$0.16 & 0.34$\pm$0.17 & 0.34$\pm$0.12 \\
\hline
\end{tabular}

\tablefoot{
\tablefoottext{a}{Background-subtracted column density.}
\tablefoottext{b}{The fit error in width $\sigma_{\rm w}$ was used for weighting data points with 1/$\sigma_{\rm w}^2$.}
}
\end{table*}

Shown in Fig.~\ref{fig_cd}\,a, compact 70\,$\mu$m sources that are protostar candidates cluster along the DR21 ridge oriented north-south, with the most prominent peak being DR21 itself.
{{Supposedly lying at the same distance,} they show a strong northward decrease in luminosity.}
Many filamentary streamers from the ridge to e.g., the north-west and west, are present, most of which were also detected in the mid-infrared with \emph{Spitzer} \citep{marston2004a,hora2009a}.
They mainly correspond to low column density structures (Fig.~\ref{fig_cd}\,b), but several {filaments} are prominent.
The extent of the DR21 ridge can be roughly defined by the N$_{\rm H_2}$$= $$10^{23}$\,cm$^{-2}$ contour enclosing an area of 2.3\,pc$^2$. Notably, the {``sub-filaments''} of \citet{schneider2010a} labeled F1 and F3 are recovered, both being resolved into nearly parallel northern and southern components that join close to the ridge.
The northern part of the DR21 ridge shows two extensions in column density (``rabbit ears'') to the north and north-west (F1) which coincide with the coldest regions where the dust temperature drops as low as 14\,K  (Fig.~\ref{fig_cd}\,c).
The {filaments} are visible in the dust temperature map as structures of lower central temperature relative to the background of $\sim$19\,K.
The background column density level is $\sim$$10^{22}$\,cm$^{-2}$ with a standard deviation of $\sim$$10^{21}$\,cm$^{-2}$ mainly due to cirrus structure.

In the first step to characterise the DR21 {filaments}, we applied the \emph{DisPerSE} software \citep{sousbie2011a} on the column density map to identify the filament crests \citep[cf.][]{arzoumanian2011a,hill2011a} using a persistence threshold of $3\times10^{21}$\,cm$^{-2}$.
{Unlike for these works, possible line-of-sight confusion needed to be addressed in our analysis due to the low-density foreground cloud Cygnus Rift and the close-by W75N cloud component \citep[e.g.,][]{schneider2006a}.
Also the strong DR21 outflow \citep[e.g.,][]{richardson1986a,white2010a} is avoided.}
Subsequent studies will analyse the combined column density, temperature, and kinematical data for a larger region, but here we conservatively limited this study to {filaments} that
(1) have main velocities in ${}^{13}$CO coherent with the ridge and no strong secondary velocity components\footnote{We used the  \citet{schneider2010a} FCRAO ${}^{13}$CO\,(1-0) data and excluded filament segments showing additional velocity components above the 20\% level of the primary peak.},
and (2) have crest column densities above $1.5\times10^{22}$\,cm$^{-2}$; {the resulting selection is} shown in Fig.~\ref{fig_cd}.
At positions along their crests spaced by half the beam FWHM we extracted perpendicular column density profiles extending over 2.5\,pc in each direction, {wide enough so that the outer profiles are sufficiently flat and trace the background}.
For background column densities, we adopted the profile minima that were median-smoothed over 2.5 beam FWHM to lower the influence of neighbouring structures.

\begin{figure}
\centering
\includegraphics[width=8.8cm]{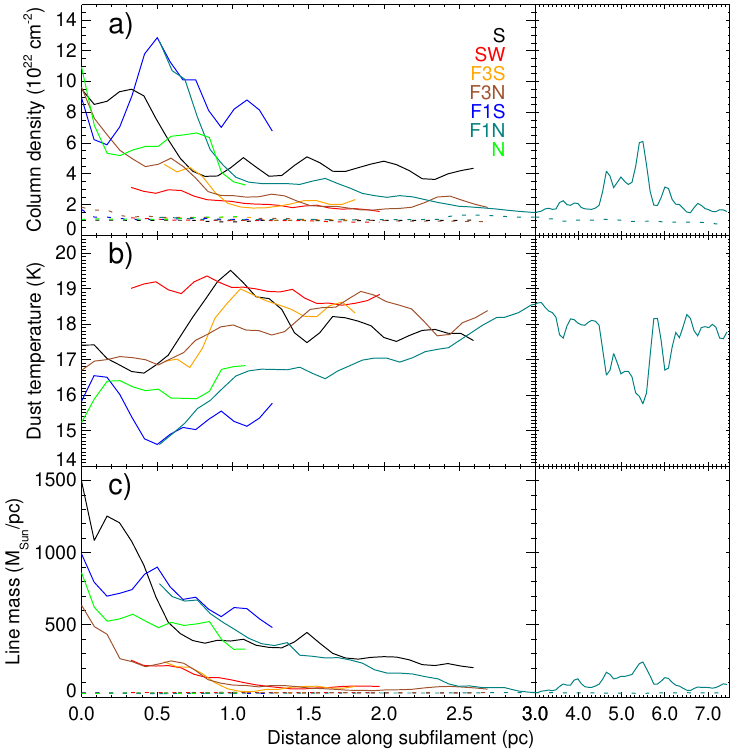}
\caption{DR21 {filament} properties over distance along their crest from the DR21 ridge. The distance scale is compressed beyond 3\,pc.
{a)} Crest column density {(not background-subtracted)}, dashed curves show the background column density,
{b)} Crest dust temperature,
{c)} Mass per unit length, dashed lines gives the critical value.
}
\label{fig_fil_cdtempwidthlm}
\end{figure}

Crest column density and dust temperature are anti-correlated {as anticipated} for all {filaments} with the possible exception of SW (Fig.~\ref{fig_fil_cdtempwidthlm}\,a, b).
Contrary to the others, the latter shows slightly decreasing central temperature outward from the DR21 ridge.
This {filament} also shows the weakest outward column density decline.
In general, lower temperatures are expected for structures of higher column density due to the increased shielding from the interstellar radiation field.
However, the SW {filament} is running towards the warmer area around DR21 (Fig.~\ref{fig_cd}\,a, c) {and possibly into the outflow cavity}.
Its inverted density-temperature relation could thus be an effect of heating by DR21.

The {perpendicular} {filament} column density profiles generally show a flattened central part and non-Gaussian, roughly power-law wings.
To examine the central width we performed a weighted fit of a Gaussian curve plus a constant level {\citep[cf.][]{arzoumanian2011a}}.
As weights a Gaussian curve with {0.17\,pc FWHM} was used to ensure the reproduction of the central flattening.
The central width {of each profile} was derived as deconvolved fitted Gaussian FWHM.
{The weighted mean values of the profile central widths lie between} {0.26} and 0.34\,pc (Table~\ref{tab_filaments}).
{These values are beyond the range around the typical $\sim$0.1\,pc found by \citet{arzoumanian2011a}, probably only partially caused by the lower spatial resolution.
However, no correlation between central width and crest column density is present, neither for profiles nor for the mean values.
Better statistics in the high-column density regime is needed to clarify a possible dependency.}

To constrain the masses of the {filaments} we estimated an outer radius for each profile integration.
A threshold per profile was derived by building the cumulative integral over radial distance from the crest and determining the first minimum of its derivative.
For each {filament} we adopted its median threshold as integration radius, ranging from 0.24 to 0.58\,pc (Table~\ref{tab_filaments}).

In theory, the thermal stability of filaments is to first order determined by the critical mass per unit length\footnote{$M_{line}^{crit} = 2c_s^2/G$ with the isothermal sound speed $c_s=\sqrt{R{\rm T}_{\rm d}/\mu_{\rm H_2}}$ and T$_{\rm d}$ at the crest position, resulting in $M_{line}^{crit} \approx$30\,M$_\odot$/pc.} \citep[e.g.,][]{inutsuka1992a}.
Observations indicate that this is a proxy for the general stability \citep{andre2010a}.
Most {filament} segments are thermally supercritical (Fig.~\ref{fig_fil_cdtempwidthlm}\,c).
The outer segments (beyond 1\,pc) of the SW, F3N, and F3S {filaments} are only supercritical to within a factor of 3, and the longest, F1N {filament} shows a marginally subcritical segment.
For thermally supercritical filaments, fragmentation and local spherical collapse occur faster than global collapse in the absence of other stabilising mechanisms \citep{pon2011a,toala2012a}.
In this case we thus expect the DR21 {filaments} to form cores and protostars.
The S {filament} shows several local maxima of crest column density seperated by about 1\,pc (Fig.~\ref{fig_fil_cdtempwidthlm}\,a).
Less regular and prominent, several peaks are also seen for the other {filaments}.
At 70\,$\mu$m there is one extended source towards the S {filament} (Figs.~\ref{fig_cd}\,a, \ref{fig_subfilaments_70_250_cd}\,a).
Compact emission from protostar candidates is present towards the junction point of F3S and F3N, and towards the F1N and N {filaments}.
The 250\,$\mu$m map shows additional compact starless/prestellar core candidates towards all {filaments}, one of them identified as dense core by \citet{motte2007a} (see Apps.~\ref{app_subfil}, \ref{app_maps}).
This confirms that core and star formation is ongoing within the thermally supercritical {filaments}.

\begin{figure}
\centering
\begin{subfigure}[b]{4.4cm}
\centering
\includegraphics[width=4.4cm]{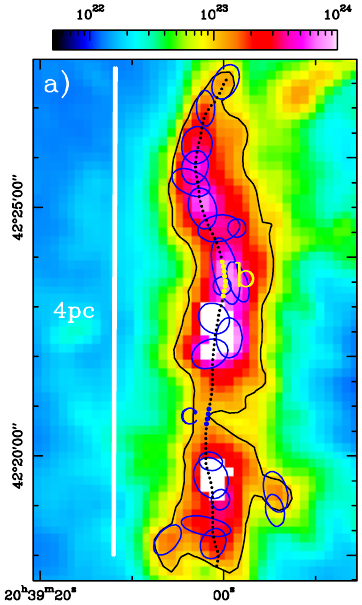}
\end{subfigure}
\begin{subfigure}[b]{4.4cm}
\centering
\includegraphics[width=4.4cm]{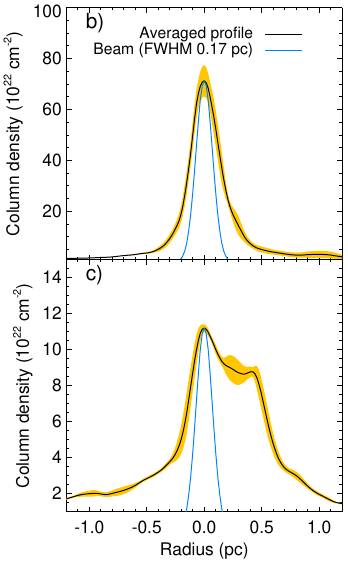}
\end{subfigure}
\caption{{a)} Column density map (in cm$^{-2}$) of the DR21 ridge with the crest as black dots and the black N$_{\rm H_2}=10^{23}$\,cm$^{-2}$ contour.
{Note that white pixels are undefined due to saturation.}
Blue ellipses give the position and extent (FWHM $=$ semiaxes) of the \citet{motte2007a} dense cores.
{b), c)} Radial column density profiles towards the crest locations b and c marked in yellow/blue in a). The radial profiles are averaged over the marked crest dots.
The yellow areas show the $\pm$1\,$\sigma$ ranges.}
\label{fig_ridge}
\end{figure}

\section{The DR21 ridge -- an intersection of filaments}
\label{sec_ridgestructure}

The DR21 ridge hosts 22 dense cores and $\sim$\,$1/4$ of the massive ones in the \citet{motte2007a} sample of the whole cloud complex, and thus represents the probable region to host the most massive forming star cluster in Cygnus~X.
We used their MAMBO 1.2\,mm map to interpolate the missing pixels in the column density map and derived the mean ridge properties using a crest of 4\,pc length following its main peaks (Fig.~\ref{fig_ridge}\,a, Table~\ref{tab_filaments}).
If it were regarded as an individual filament, it would be highly unstable, in accord with the global collapse signature observed by \citet{schneider2010a} and the dense core detections (Fig.~\ref{fig_ridge}\,a).

The chain of cores dominates the structure of the ridge except between DR21 and DR21(OH), where a crest column density dip is present (location c in Fig.~\ref{fig_ridge}).
Less clear than the northern branching, a possible additional filamentary extension coinciding with a core branches off to the south-east at the southern end of the ridge {(at the very bottom of Fig.~\ref{fig_ridge}\,a)}.
As illustrated in Fig.~\ref{fig_ridge}\,b for {location b}, corresponding to the dense core N40, the column density profile of the ridge generally shows narrow peaks and smoothly decreasing wings towards the cores.
However, the profile towards location c shows at least one additional peak of 9$\times$10$^{22}$\,cm$^{-2}$ at 0.4\,pc (Fig.~\ref{fig_ridge}\,c), indicating that there the ridge consists of more than one individual filament.
Considered together with the branching of the northern ridge into the ``rabbit ears'' (the F1 and N {filaments}) and the possible southern branching, this finding suggests that the DR21 ridge is a complex intersection of several individual filaments.
The dominating filament along the crest shows a central width of $\sim$0.34\,pc, not significantly larger than the central widths of the {filaments} (Table~\ref{tab_filaments}).

\section{Star formation in the DR21 ridge}

\citet{schneider2010a} present velocity gradients that suggest accretion of the {filaments} onto the DR21 ridge.
Dust temperature dips are seen towards the {filament}-ridge connections (Fig.~\ref{fig_fil_cdtempwidthlm}\,b), indicating that inflowing material has cooled at these locations.
At present, the total mass in the selected {filaments} is about 1/3 of the ridge mass.
It thus appears that the mass assembly process of the DR21 ridge is in a late stage, possibly driven by the gravitational potential of the ridge itself, and specifically the build-up of massive cores in the DR21/DR21(OH) clumps could have been supported by continuous accretion from the S/SW and F3S/F3N {filaments}, respectively.
The {filaments} are for the most part gravitationally unstable and forming cores and protostars, in contrast to the striations and {``sub-filaments''} in Taurus \citep[][Palmeirim et al. in prep.]{goldsmith2008a} or Aquila \citep{andre2010a}.
The high masses of the cores in the ridge could thus not only be due to the {filament} flows, but also due to the merging with fragment cores of the {filaments} that form dense, small-scale convergent flows.
For DR21(OH), \citet{csengeri2011b} showed possible fragmentation of the inflowing material.
High-mass star formation involving a core merger process was also suggested by observations of NGC\,2264-C \citep{peretto2006a}.

The \emph{Herschel} observations emphasize the evolutionary gradient along the ridge: Beyond DR21, the 70\,$\mu$m luminosity of protostars strongly decreases northward, and the dust temperature shows lowest temperatures towards the northern part.
The substructure of the DR21 ridge suggests that it was formed by the merging of individual narrow, intersecting filaments.
The pronounced elongation and branching into {filaments} F1 and N with the highest mass per unit length indicate that the major components may be roughly parallel filaments oriented north-south.
At present, the merging could have advanced furthest in the southern part and less in the north, where the two components appear separated.
Extrapolating this scenario, we expect that the northern {filaments} will lead to the assembly of one or more further massive clumps.
This study suggests that high-mass star-forming ridges could be second-generation cloud structures formed via dynamical merging of gravitationally unstable filaments.

\bibliographystyle{aa} \bibliography{/Users/mhennema/work/literature/library} 
\begin{acknowledgements}
SPIRE has been developed by a consortium of institutes led
by Cardiff Univ. (UK) and including: Univ. Lethbridge (Canada);
NAOC (China); CEA, LAM (France); IFSI, Univ. Padua (Italy);
IAC (Spain); Stockholm Observatory (Sweden); Imperial College
London, RAL, UCL-MSSL, UKATC, Univ. Sussex (UK); and Caltech,
JPL, NHSC, Univ. Colorado (USA). This development has been
supported by national funding agencies: CSA (Canada); NAOC
(China); CEA, CNES, CNRS (France); ASI (Italy); MCINN (Spain);
SNSB (Sweden); STFC, UKSA (UK); and NASA (USA).
PACS has been developed by a consortium of institutes led by MPE (Germany) and including UVIE (Austria); KU Leuven, CSL, IMEC (Belgium); CEA, LAM (France); MPIA (Germany); INAF-IFSI/OAA/OAP/OAT, LENS, SISSA (Italy); IAC (Spain). This development has been supported by the funding agencies BMVIT (Austria), ESA-PRODEX (Belgium), CEA/CNES (France), DLR (Germany), ASI/INAF (Italy), and CICYT/MCYT (Spain).
This work was supported by the ANR (\emph{Agence Nationale pour la Recherche})
project ``PROBeS" (ANR-08-BLAN-0241).
\end{acknowledgements}

\appendix

\section{The S and F1N {filaments} in detail}
\label{app_subfil}

\begin{figure*}
\centering
\includegraphics{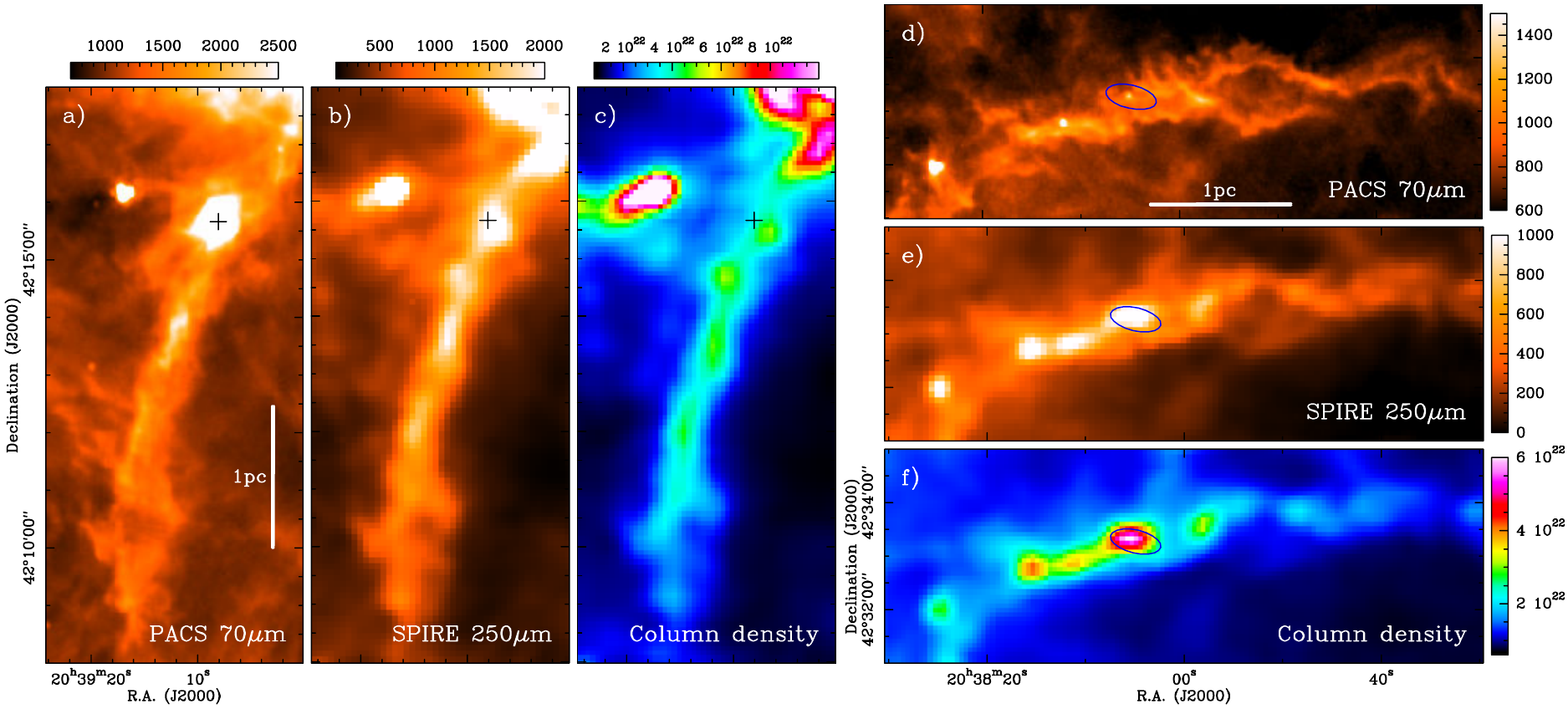}
\caption{Detailed view of {filament} S (panels a, b, c) and F1N (d, e, f).
The 70 and 250\,$\mu$m units are MJy/sr, column density unit is cm$^{-2}$.
The black cross in panel a) indicates an extended 70\,$\mu$m source.
The blue ellipse in panels d), e), f) gives the position and extent (FWHM $=$ semiaxes) of the dense core N23 of \citet{motte2007a}.}
\label{fig_subfilaments_70_250_cd}
\end{figure*}

Figure~\ref{fig_subfilaments_70_250_cd} shows a zoom on the S and F1N filaments.
Several compact, more or less elongated cores are seen in the 250\,$\mu$m and column density map.
The marked dense core N23 has an estimated mass of 13\,M$_\odot$.

\Online

\section{\emph{Herschel} maps of DR21 and Cygnus X North}
\label{app_maps}

\begin{figure*}
\centering
\includegraphics[width=16cm]{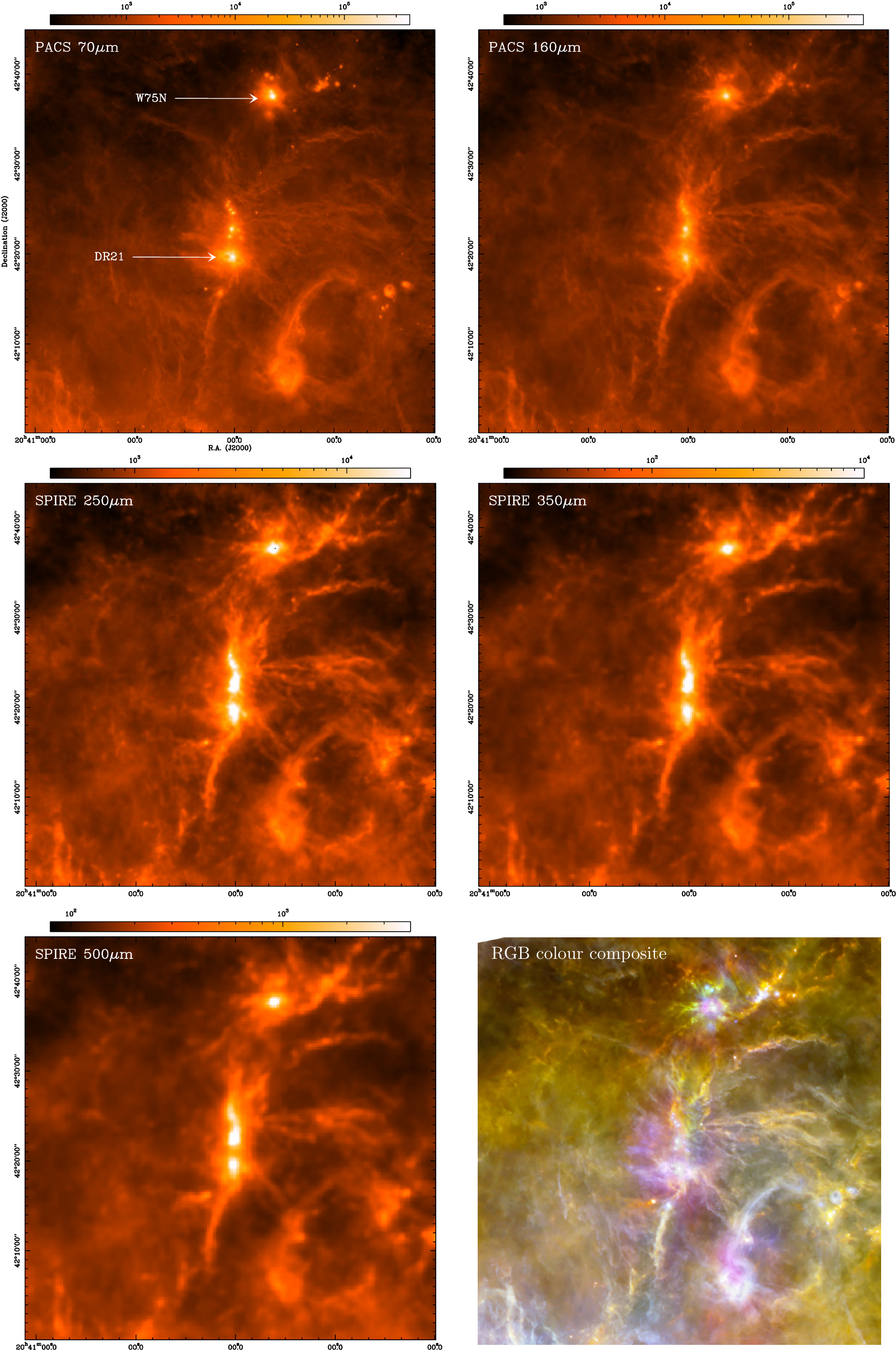}
\caption{\emph{Herschel} maps of the DR21 environment in Cygnus~X North obtained by HOBYS.
Map units are MJy/sr.
The last panel shows a RGB colour composite image using the SPIRE 250\,$\mu$m (red), PACS 160\,$\mu$m (green), and PACS 70\,$\mu$m (blue) maps.}
\label{fig_maps}
\end{figure*}

Figure~\ref{fig_maps} shows the DR21 environment in the \emph{Herschel} maps of the HOBYS Cygnus~X North observations together with a RGB colour composite image.

\end{document}